\documentclass[sigconf]{aamas} 

\usepackage{balance} 

\setcopyright{ifaamas}
\acmConference[AAMAS '23]{Proc.\@ of the 22nd International Conference
on Autonomous Agents and Multiagent Systems (AAMAS 2023)}{May 29 -- June 2, 2023}
{London, United Kingdom}{A.~Ricci, W.~Yeoh, N.~Agmon, B.~An (eds.)}
\copyrightyear{2023}
\acmYear{2023}
\acmDOI{}
\acmPrice{}
\acmISBN{}

\acmSubmissionID{297}

\title{Strategic Planning for Flexible Agent Availability in Large Taxi Fleets}

\author{Rajiv Ranjan Kumar}
\affiliation{
	\institution{School of Computing and Information Systems\\Singapore Management University}
    \country{Singapore}}
\email{rajivrk.2017@phdcs.smu.edu.sg}

\author{Pradeep Varakantham}
\affiliation{
	\institution{School of Computing and Information Systems\\Singapore Management University}
    \country{Singapore}}
\email{pradeepv@smu.edu.sg}

\author{Shih-Fen Cheng}
\affiliation{
	\institution{School of Computing and Information Systems\\Singapore Management University}
    \country{Singapore}}
\email{sfcheng@smu.edu.sg}


\newcommand{\BibTeX}{\rm B\kern-.05em{\sc i\kern-.025em b}\kern-.08em\TeX}

\usepackage{bibentry}
    
\usepackage{algorithm}
\usepackage{algpseudocode}
\usepackage{color, colortbl}

\usepackage{subfig}

\usepackage{amsmath}

\DeclareMathOperator*{\argmin}{argmin}
\definecolor{Grey}{rgb}{0.5,0.5,0.5}
\definecolor{LightCyan}{rgb}{0.88,1,1}
\definecolor{Gray}{gray}{0.85}
\newcolumntype{a}{>{\columncolor{Gray}}c}

\newtheorem{exple}{Example}

\newcommand{\squishlist}{
	\begin{list}{$\bullet$}
		{ \setlength{\itemsep}{0pt}
			\setlength{\parsep}{2pt}
			\setlength{\topsep}{2pt}
			\setlength{\partopsep}{0pt}
			\setlength{\leftmargin}{1.5em}
			\setlength{\labelwidth}{1em}
			\setlength{\labelsep}{0.5em} } }
	
	\newcommand{\squishend}{
\end{list}  }

\begin{abstract}
In large-scale multi-agent systems like taxi fleets, individual agents (taxi drivers) are self-interested (maximizing their own profits) and this can introduce inefficiencies in the system. One such inefficiency is with regard to the "required" availability of taxis at different time periods during the day. Since a taxi driver can work for a limited number of hours in a day (e.g., 8-10 hours in a city like Singapore), there is a need to optimize the specific hours, so as to maximize individual as well as social welfare.  Technically, this corresponds to solving a large-scale multi-stage selfish routing game with transition uncertainty. Existing work in addressing this problem is either unable to handle ``driver" constraints (e.g., breaks during work hours) or not scalable. To that end, we provide a novel mechanism that builds on replicator dynamics through ideas from behavior cloning. We demonstrate that our methods provide significantly better policies than the existing approach in terms of improving individual agent revenue and overall agent availability.
\end{abstract}

\begin{document}

\begin{abstract}

In large scale multi-agent systems like taxi fleets, individual agents (taxi drivers) are self interested (maximizing their own profits) and this can introduce inefficiencies in the system. One such inefficiency is with regards to the "required" availability of taxis at different time periods during the day. Since a taxi driver can work for limited number of hours in a day (e.g., 8-10 hours in a city like Singapore), there is a need to optimize the specific hours, so as to maximize individual as well as social welfare.  Technically, this corresponds to solving a large scale multi-stage selfish routing game with transition uncertainty. Existing work in addressing this problem is either unable to handle ``driver" constraints (e.g., breaks during work hours) or not scalable. To that end, we provide a novel mechanism that builds on replicator dynamics through ideas from behavior cloning. We demonstrate that our methods provide significantly better policies than the existing approach in terms of improving individual agent revenue and overall agent availability. 
\end{abstract}

\maketitle

\section{Introduction}
Aggregation systems offer time sensitive tasks/services (require workers to respond within a minute). Notable examples of such platforms include transport (e.g., Uber or Grab or Gojek), food delivery (e.g., Foodpanda or DoorDash), and grocery delivery (e.g., Amazon Fresh or Instacart). Workers on these platforms have to constantly make both spatial and temporal decisions: the ``spatial'' decision on ``where'' to work and the ``temporal'' decision on ``when'' to work. These decisions are challenging since the performance of workers depends not just on their understandings of the demands, but also on when and where other workers choose to work. So, in essence, there is a need to study strategic and operational decision problems together. 

In the literature, such decision problems are typically formulated using game theory and solved using equilibrium-seeking algorithms (since an individual has to best respond to all others' decisions, which are best responses to our best responses as well, \emph{ad infinitum}). However, existing research~\citep{varakantham2012decision,gan2017game} consider either spatial decisions or temporal decisions but not both. Even in the one recent work where both spatial or temporal decisions are considered~\citep{kumar2021adaptive}, there are rigid assumptions on workers' work arrangement (e.g., although a worker can choose when to start working, once work starts, this worker has to work consecutively for a fixed number of hours). Furthermore, the equilibrium computation approaches that are employed are computationally expensive (with 3 days to solve one instance) and do not scale easily to real problems of interest. 

To that end, we  develop scalable approaches that consider both spatial and temporal decisions for workers (henceforth referred to as agents), while allowing for flexibility in agent  availability. Agents are profit-maximizing and their strategic decisions include: 1) pick the time period to enter the system and start working; 2) once in the system, choose the regions to move to at different times; and 3) while in the system, an agent can choose to take a temporary break of arbitrary length, subject to agents' total working hours.

Due to the scale of problems at hand with thousands of agents, existing methods have exploited homogeneity and anonymity (lack of identity) to compute symmetric equilibrium solutions~\citep{kumar2021adaptive,varakantham2012decision}. However, such approaches are not scalable and can take 3 days to run a single instance. To achieve scalability, we propose a simulation based mechanism  based on the ``replicator dynamics framework''~\citep{weibull1997evolutionary} to compute approximate symmetric equilibrium solutions.  Intuitively, at each iteration of our new method, the current symmetric policy is simulated for all the agents and the top ``k" (where k is around 500) trajectories (in terms of revenue generated) are employed to compute a new symmetric policy. Since we consider a finite ``k" it is feasible that some states are not covered by the trajectories. Therefore, we also explore multiple data imputation mechanisms to complete the policy.

Our paper makes the following key contributions:
\begin{enumerate}
	\item We introduce a game-theoretic model that allows agents to choose their working schedule in a flexible manner. After picking a time to enter the system and begin working, agents can choose when to take breaks and where to work.
	\item We propose a simulation-based best response method for computing symmetric equilibrium at scale.
	\item To handle partial policy samples obtained with our  simulation method, we propose various data imputation techniques and  show the utility of these methods in improving solution quality efficiently.
	\item We demonstrate the scalability of our approach by solving problem instances that contain more than 20,000 agents in 10 minutes. 
\end{enumerate}

\section{Related Work}

Past work has employed approaches for operational optimization of taxi fleets  using heuristic approaches~\citep{maciejewski2015large,maciejewski2016assignment}, bipartite matching~\citep{vazifeh2018addressing,horl2019fleet}, multi-objective optimization models~\citep{miao2016taxi} and deep-learning based methods~\citep{oda2018movi,gao2018optimize,xu2018large}). There are multiple key differences with these line of works. First, above approaches assume a centralized optimization (by aggregation companies like Uber) and individual taxis are assumed to follow the inputs provided by the central entity. Second, the focus in the above approaches is typically on solving the operational decision problem of which zone to move to at different points of the day if there is no customer on board. In contrast, we focus on the combination of strategic (when to start shift) and operational (where to move to) problems with decentralized optimization (where individual taxis maximize their own revenue). 

\citet{saisubramanian2015risk} used MIP to optimize the ambulance allocation in the emergency department
setting. \citet{lowalekar2017online} proposed online repositioning solution for bike sharing systems. For taxi fleet optimization there has been some recent work on decentralized approaches~\citep{yuan2012t,zhang2016framework,moreira2012predictive,qu2014cost,varakantham2012decision}, however the emphasis there has been on the operational decision problem  and not on the strategic decision problem (optimizing starting time of shift and breaks), which is a main focus in this paper. 

 The closest literature to our contributions in the paper are  \citep{gan2017game,kumar2021adaptive}. While  \citet{gan2017game} considers driver's operating hours constraints explicitly to optimize constraints related to operation of taxi drivers, \citet{kumar2021adaptive} incorporates starting time period as part of agent’s strategies (not constraints) to provide advice on when to start the shift. Such approaches are not scalable and work with rigid assumptions on taxi's work arrangement. In comparison, our work allows for flexibility in agent availability by optimizing and allowing breaks in operation. Furthermore, our approach is significantly more scalable (10 minutes vs 3 days) due to the simulation based best response framework employed to compute approximate symmetric equilibrium solutions. 

\section{Flexible Taxi Fleets}

In many cities where taxis play an important role in providing point-to-point transport, it is of critical importance for drivers to decide how to position themselves when idling, so as to maximize their revenues. The driver's decision on where to move to is strategic in nature, as drivers are competing for the same pool of jobs when they are close to each other. 

In this paper, we aim to introduce another important dimension in driver's strategy space, which is when to work. Similar to the spatial movement decisions, decisions on when to work are also strategic in nature for driver, as the number of drivers per time period also has a strong impact on individual's revenue potential. Compared to a recent work by \citet{kumar2021adaptive}, which also allows drivers to pick their work starting time, our proposed model is much more flexible in that we allow drivers to freely leave and return to the service platform (as opposed to having to pick a starting time and work continuously for a fixed amount of time). Our proposed model is thus much more realistic and expressive.



We formally define our model as Flexible Selfish Routing under Uncertainty (FSRU), which allows workers to freely take break during their service.  Our FSRU framework is a special case of the generic stochastic game model~\citep{shapley1953stochastic, neyman2003stochastic, mertens1981stochastic}: while the transition and reward functions for an agent in our model are dependent only on the aggregate distributions of other agent states, they can be dependent on specific state and action of every other agent in general stochastic games. FSRU extends on the Selfish Routing under Transition Uncertainty (SRT) model~\citep{kumar2021adaptive}, by allowing for flexible operations of individual agents. FSRU represents problems with selfish agents and hence is different to cooperative models such as Decentralized POMDPs~\citep{bernstein2002complexity}. A FSRU instance is characterized by the tuple:
$$\Big< {\mathcal P}, {\mathcal S}, {\mathcal A}, {\mathcal T}, \{{\mathcal R}_{\tau}\}_{\tau \in \Gamma}, B, \boldsymbol{d}^{0}, \delta, H\Big>$$
where ${\mathcal P}$ is set of agents, i.e., the set of taxis. 
${\mathcal S}$ is the set of zones a taxi could move to, ${\mathcal A}$ is the set of zones to which a taxi driver wishes to move. There is a special ``Sink" state, $s_{sink}$, that is used to represent a taxi driver on a break and an action $a_{sink}$ that represents taking a break and will move the state to $s_{sink}$ state with probability 1. If a taxi in $s_{sink}$ state takes any action, action will succeed with probability 1. 

${\mathcal T}$ models the transition function of every agent and more specifically, ${\mathcal T}^t(s,a,s',\boldsymbol{d})$ represents the probability that an agent in state $s \in {\mathcal S}$ after taking action $a \in {\mathcal A}$ would transition to state $s'$, when the state distribution of all (active) agents is $\boldsymbol{d}$ at time $t$. In case of taxi fleet, the transition function, ${\mathcal T}$, depends on the involuntary movements between zones. The involuntary movement between any two zones $i$ and $j$ at a time step $t$ is determined by the number of customers (customer flow), $fl^t(i,j)$ moving between those zones. Thus, demand for taxis in a zone $i$ at time step $t$ is given by $\sum_j fl^t(i,j)$. We assume that the zones are small enough that if the number of taxis in a particular zone during a time period is fewer than the demand, then all taxis will be hired. If the number of taxis are more than the demand, a fraction of the taxis (equal to demand) will be hired and each taxi will have equal probability of getting hired.  Equation~\eqref{eqn:phi} provides the expression for computing the transition probabilities between states.  
\begin{flalign}
		&{\mathcal T}^t(s,a,s',\boldsymbol{d}) = \nonumber \\
		&\left\{
		\begin{array}{ll} 
			1, & \hspace{-1in}\text{ if } (a=a_{sink}\;\&\; s' = s_{sink}) \text{ or } (s=s_{sink}\;\&\; a = s') \\
			0, & \hspace{-1in}\text{ if } (a=a_{sink}\;\&\; s' \ne s_{sink}) \text{ or } (s=s_{sink}\;\&\; a \ne s')\\
			\frac{{fl}^t(s,s')}{\sum_{\hat{s}} {fl}^t(s,\hat{s})} & \;\text{if } s \ne s_{sink}, a \ne a_{sink} \text{ and } (\textbf{C1})  \vspace{0.05in}\\
				\frac{{fl}^t(s,s')}{d_s} & \; \text{if } s \ne s_{sink}, a \ne a_{sink} \text{ and } (\textbf{C2}) \vspace{0.05in}\\
				1 - \frac{\sum_{\hat{s} \neq s'}{fl}^t(s,\hat{s})}{d_s} & \;  \text{if } s \ne s_{sink}, a \ne a_{sink} \text{ and } (\textbf{C3})
			\end{array}
			\right.
			\label{eqn:phi}
\end{flalign}

The exact transition probabilities depend on normalized demands to other zones from the current zone (\textbf{C1} represents this case; \textbf{C1}: $\sum_{\hat{s}} {fl}^t(s,\hat{s}) \geq d_s$). On the other hand, if the number of taxis is more than the number of customers in a zone, the transition probabilities should depend on whether the action (intended zone) coincides with the destination zone (\textbf{C2} and \textbf{C3} represent these two cases). The condition \textbf{C2} is only possible when the taxi agent is hired by a customer heading towards any $s'$ that is not $a$ (\textbf{C2}: $\sum_{\hat{s}} {fl}^t(s,\hat{s}) < d_s \text{ and } a \neq s'$). For condition \textbf{C3}, the taxi agent can either be free or hired by a customer heading towards the agent's intended zone (\textbf{C3}: $\sum_{\hat{s}} {fl}^t(s,\hat{s}) < d_s \text{ and } a = s'$).




${\mathcal R}_{\tau}^t(s,a,s',\boldsymbol{d})$ is the reward obtained by an agent of type $\tau$ at time t when in state $s$, taking action $a$ and moving to state $s'$ when the state distribution of other agents is $\boldsymbol{d}$ at time $t$. 	Similar to the transition probabilities, the reward function for all active agents at time t. We will refer to the cost (of moving) and revenue (of moving a customer) from state $s$ to $s'$ at time $t$ as $c^t(s,s')$ and $r^t(s,s')$ respectively. 
${\mathcal R}^t(\cdot)$ is defined differently under these three conditions:

	\begin{flalign}
		\hspace{-0.5in}{\mathcal R}^t(s,a,\boldsymbol{d}) &= \left\{
		\begin{array}{ll}
			c^t(s,s_{sink})  &\hspace{-1in}\text{ if } a=a_{sink} \\
			c^t(s_{sink},s'=a) &\hspace{-1in}\text{ if } s = s_{sink}   \\
			\sum_{s'}{\mathcal T}^t(s,a,s',\boldsymbol{d}) \cdot \left(r^t(s,s') - c^t(s,s')\right)  \\ & \hspace{-2in}\;\text{if } s \ne s_{sink}, a \ne a_{sink} \text{ and } (\textbf{C1})\vspace{0.1in}\\
			\sum_{s' \neq a}{\mathcal T}^t(s,a,s',\boldsymbol{d}) \cdot \left(r^t(s,s') - c^t(s,s')\right) \\ & \hspace{-2in}\;\text{if } s \ne s_{sink}, a \ne a_{sink} \text{ and } (\textbf{C2})\vspace{0.1in}\\
			\frac{{fl}^t(s,a)}{d_s} \cdot r^t(s,a) - {\mathcal T}^t(s,a,a) \cdot c^t(s,a) \\  & \hspace{-2in}\;\text{if } s \ne s_{sink}, a \ne a_{sink} \text{ and } (\textbf{C3})
		\end{array}
		\right.
		\label{eqn:ri}
	\end{flalign}

It should be noted that taxis are hired in conditions \textbf{C1} and \textbf{C2}; therefore, the expected rewards in these two cases are the sum of expected rewards to all feasible destinations. For \textbf{C3}, cost is incurred for sure, but revenue can only be earned if the taxi is hired. Our goal in solving taxi fleet optimization problem as a FSRU is to maximize expected revenue for individual taxi drivers who are perfectly rational and follow computed policies.

$B$ represents maximum number of allowed breaks in operation. $\delta$ is the maximum number of time steps any agent can be active. Finally $H$ represents the time horizon of the decision making process. 

\emph{\textbf{The good thing about the FSRU model is that all the elements of the tuple can be computed from any real world taxi dataset, where information (i.e., source, destination, fare, cost) about customer requests is available.}}

\subsection{Example}

We show how transition probabilities and the rewards are computed for a small  problem extended from the one introduced in~\citet{kumar2021adaptive}:
\newsavebox\transitionval
\begin{lrbox}{\transitionval}
  \begin{minipage}{0.5\textwidth}
    \begin{align*}
      {\mathcal T}^t_d(s,a,s') &= 0.0  if s=s'\\
		      &= 0.5
    \end{align*}
  \end{minipage}
\end{lrbox}
\begin{exple}
Consider a map with three zones, ${\mathcal S}=\{s_0,s_1,s_2\} \cup s_{sink} $. For simplicity we set flow values to 1 across all time periods; i.e., one passenger goes from each zone to an adjacent zone in all time periods. We also set rewards and costs to be fixed at $r^t(s,s')=1$ and $c^t(s,s')=0$ for all time periods $t$ and all zones $s, s' \in {\mathcal S}$. If the distribution of taxis at a given time period $t$ is $\boldsymbol{d}^t=(1, 1, 4, 0)$, then:
\squishlist \item{The transition function ${\mathcal T}^t(s,a,s',\boldsymbol{d}^t)$ is specified by matrix $m(s)$, in which the row label represents action $a$, and the column label represents destination zone $s'$. Transition function for $s_0$ and $s_2$ are: $m(s_0) \text{ \& }  m(s_2)=$
{ \[ \hspace{-0.15in} \left( \begin{array}{cccc}
0.0 & 0.5 & 0.5 & 0.0\\
0.0 & 0.5 & 0.5 & 0.0\\
0.0 & 0.5 & 0.5 & 0.0\\
0.0 & 0.0 & 0.0 & 1.0\end{array} \right) \&
\left( \begin{array}{cccc}
0.75 & 0.25 & 0.0 & 0.0\\
0.25 & 0.75 & 0.0 & 0.0\\
0.25 & 0.25 & 0.5 & 0.0\\
0.0 & 0.0 & 0.0 & 1.0\end{array} \right) \]
}}

\item{Similarly, the reward function ${\mathcal R}^t(s,a,\boldsymbol{d})$ is specified as a matrix, in which the row label represents current zone $s$, and the column label represents action $a$:
\[  \left( \begin{array}{cccc}
1.0 & 1.0 & 1.0 & 0.0\\
1.0 & 1.0 & 1.0 & 0.0\\
0.5 & 0.5 & 0.5 & 0.0\\
0.0 & 0.0 & 0.0 & 0.0\end{array} \right)\] 
}
\squishend
Consider state $s_2$: $fl^t(s_2,s')=1$ for $s' \in \{s_0,s_1\}$ and $d^t(s_2)=4$. The transition and reward functions are computed using \eqref{eqn:phi} and \eqref{eqn:ri}. For $a=s_2$ and $s' \in \{s_0,s_1\}$, ${\mathcal T}^t(s_2,a,s',\boldsymbol{d}^t)= 1/4$ by \textbf{C2}. For $a=s_2$ and $s'=s_2$, ${\mathcal T}^t(s_2,a,s',\boldsymbol{d}^t)= 1-2/4=1/2$ by \textbf{C3}.
For $a=s_2$ and $s'=s_{sink}$, ${\mathcal T}^t(s_2,a,s',\boldsymbol{d}^t)= 0$. Rest of the transition and reward function values can be computed similarly.
\end{exple}




\section{Brute-force Approach: FP}
\begin{algorithm}[htbp!]
	\caption{Finding $\epsilon$-Equilibrium}
	\begin{algorithmic}[1]
		\State	$\pi_0$ = GetRandomPolicy(), $x_0$ = GetAgentFlow($\pi_0$,N)
		\State	i = 1, converged = false
		\While {converged = false}
		\State	$probability$ = GetAgentCountProbability($\pi_0$,N-1)
		\State	$x_1$ = GetBestResponse($probability$)
		\State	$x_1^t(s,n,b,a) = (x_0^t(s,n,b,a) \cdot i +$ \\\hspace{2cm} $x_1^t(s,n,b,a))/(i+1), \quad \forall t,s,a,n$
		\State	${\pi_1}^t(s,n,b,a)$ =  $x_1^t(s,n,b,a)/ \sum_a x_1^t(s,n,b,a),$
		 $\forall t,s,n,b,a$
		\If {$\mid \pi_0 - \pi_1 \mid  \le \epsilon_\phi$}
		\State	converged = TRUE
		\Else
		\State	$x_0 = x_1,$ \quad $\pi_0 = \pi_1$
		\EndIf
		\State	$i+1$
		\EndWhile
	\end{algorithmic}
	\label{alg:FP_New}	
\end{algorithm}
In this section, we  describe the Fictitious Play (FP~\cite{brown1951iterative}) approach for computing approximate symmetric Nash Equilibrium (NE) solutions in FSRU. 

\begin{algorithm}[!h]
	\caption{GetBestResponse(p)}
	\begin{algorithmic}
		\State $s_{S}$: sink state, $s_{\neg S}$: not sink state
		\State $a_{S}$: sink action, $a_{\neg S}$: not sink action
		\State $M$: maximum allowed operating hours 
		\State $B$: maximum allowed breaks in operation
		\begin{align}
			&\max \sum_{t,s,n,b,a,i} x^t(s,n,b,a) \cdot p^t_i(s) \cdot R^t(s, a, i+1) \\
			& \textbf{s.t}\;\; \sum_{a} x^0(s_{S},0,0,a) = 1 \\
			& \sum_{n \in (1,M), b \in (1,B), a} x^0(s_{S},n,b,a) = 0 \\			
			& \sum_{a} x^t(s_{\neg S},n,b,a) \nonumber \\ & 
			= \sum_{s',a'}
			x^{t-1}(s'_{\neg S},n-1,b,a'_{\neg S})  \cdot  p^{t-1}_i(s')  \cdot \phi^{t-1}_{i+1}(s',a,s) \nonumber \\ &  \quad  +  \sum_{s',a_{S}} x^{t-1}(s'_{\neg S},n-1,b-1,a_{S})   \cdot p^{t-1}_i(s') \nonumber \\ &  \quad \cdot \phi^{t-1}_{i+1}(s',a,s) + x^t(s_{S},n,b,a=s), \quad \forall t, n, s_{\neg S} \label{cons:flow1}\\
			&\sum_{a} x^t(s_{S},n,b+1,a) = \sum_{s'}
			x^{t-1}(s',n,b,a_{S}),  \nonumber \\ &  \qquad \qquad \qquad \qquad \qquad \qquad   \forall t \in (1,T),n \\
			&\sum_{t,s,a_{\neg S}} x^t(s,n=M,b,a_{\neg S}) = 0 \\
			&\sum_{t,s,a_{\neg S}} x^t(s,n,b=B,a_{\neg S}) = 0 
		\end{align}
	\end{algorithmic}
	\label{alg:bestResponseMDP_New}
\end{algorithm}
Fictitious play is an iterative approach, where an agent plays best response against aggregate policy (aggregated over all the previous iterations) of all other agents at each iteration. Algorithm~\ref{alg:FP_New} provides this fictitious play algorithm for FSRU building on work by \citet{varakantham2012decision}, where the best response for an agent (line 7) is computed by using Algorithm~\ref{alg:bestResponseMDP_New}. 

Towards achieving a flexible shift optimization, we embed the policy with shift information of agents, i.e., policy for any agent is defined as, $\pi^t(s, n, b, a)$, where, t is time horizon, s is state, a is action, n is number of hours served by the agents and b is breaks taken by the agent.
$$\sum_{a} \pi^t(s,n,b,a) = 1 \qquad \forall t,s,n,b$$
By embedding shift information inside policy we allow for agents operating hours to be split into multiple parts. We provide a FP-based mechanism (Algorithm~\ref{alg:FP_New}) to compute an approximate equilibrium solution if the approach converges. This FP process at each iteration computes best response (line 5) against aggregate policy of all other agents (line 4). It uses Algorithm~\ref{alg:bestResponseMDP_New} for computing best response and is based on the dual LP formulation for solving MDPs. It is more complicated than a basic MDP, because of the flexible operating hours constraint (breaks). 

There are two key challenges. First, policy space has two new dimensions [$\pi^t(s,a) \rightarrow \pi^t(s,n,b,a)$], which increases the complexity of fictitious play process significantly. Second, in best response computation, initial distribution is not known and agents can leave (take a break) and join the simulation multiple times (we do not know when and where agent starts its operation and agent can start multiple times after taking a break).

To handle issue on the initial distribution, we introduce a dummy node for agents. At the beginning of the process, i.e., at time step 0, all agents start their operation from a node decided by policy, to avoid initial distribution issue (where initial distribution in unknown) all agents are assumed to be at dummy node and policy will decide to move the taxi to any other node in the next time step. Agents can move from and to the dummy node. This ensures that the decision problem is no longer the initial distribution, but the action to take in each state. These changes are incorporated into Algorithm~\ref{alg:bestResponseMDP_New} for best response computation.



\section{Best Response via Simulation} 
For one instance, the FP process requires many iterations to converge and in each iteration it has to go through a time consuming best response computation. Due to this, it can take weeks to compute a solution. More importantly, the solutions computed are not effective if the process is stopped early. In addition, we also have to run multiple instances with different input data values to get the best strategy.

Therefore, for efficient and effective computation, we replace the Algorithm~\ref{alg:bestResponseMDP_New} with a simulation-based best response method in  Algorithm~\ref{alg:sim_bestResponse}. Simulation-based best response takes policy as input, as the policy is simulated by all agents to identify top trajectories. This is unlike in the FP approach of the previous section, where agent count probability (line 6 of Algorithm~\ref{alg:bestResponseMDP_New}) is the input. 

In, Algorithm~\ref{alg:sim_bestResponse}, we first compute
revenue weighted occupancy measure over k-best response paths (line 1) by using Algorithm~\ref{alg:sim_bestResponsePath} and form the path matrix (line 2) which contains no information (marked as NAN or NULL) for non-visited states-actions and the weighted occupancy measure for visited states-actions. 
Algorithm~\ref{alg:sim_bestResponsePath} computes the weighted best response path for top-k agents. We first simulate agent policies (in line 2) and identify the paths (or trajectories) taken by top-k agents (lines 3-4) in terms of revenue. We then compute one path from top-k paths, weighted according to their revenues (line 5). 

\begin{algorithm}[!h]
	\caption{GetBestResponsePath-Sim-Topk($\pi$)}
	\begin{algorithmic}[1]
		\State P: set of agent paths, R: set of agent revenue
		\State P, R = SimulateAgents($\pi$)
		\State P, R = SortPathAndRevenue\_RevenueWise(P, R)
		\State $TopP_k$, $TopR_k$ = GetTopKPathsAndRevenue(P, R)
		\State weightedPath = getWeightedPath($TopP_k$, $TopR_k$)
		\State \textbf{return} weightedPath
	\end{algorithmic}
	\label{alg:sim_bestResponsePath}
\end{algorithm}
\begin{algorithm}[!h]
	\caption{GetBestResponsePolicy-Sim-Topk($\pi$) }
	\begin{algorithmic}[1]
		\State weightedPath = GetBestResponsePath-Sim-Topk($\pi$)
		\State pathMatrix = getPathMatrix(weightedPath)
		\State pathMatrix = Impute(pathMatrix)
		\State $\pi$ = GetPolicy(pathMatrix)
		\State \textbf{return} $\pi$
	\end{algorithmic}
	\label{alg:sim_bestResponse}
\end{algorithm} 

Simulation-based best response computation developed here is built on the idea of ``Replicator Dynamics''~\citep{weibull1997evolutionary} from the ``Evolutionary Game Theory''. Replicator dynamics has two key properties. First, if a trajectory/path is absent from the population, then it must always have been absent and will always be absent at any time in the future. Second, if a trajectory/path is present in the population, then it must always have been present and will always remain present at any time in the future. 


When we compute best response using simulation, the best response is bounded by the input policy, e.g., if in a state $s$, the probability for taking an action $a$ is 0, then the best response agent will never be able to take action $a$ in state $s$, and this may lead to sub optimal best response because agent is not able to explore all possibilities. 
To ensure best response computation is able to explore all possibilities before making its decision, instead of using a random starting policy (as done in line 1 of Algorithm~\ref{alg:FP_New}), we design a uniform starting policy that assigns equal importance along both spatial as well as temporal dimensions (given the constraints on maximum hours to serve and maximum number of breaks) of state space. 


If we only consider the best path, only a small part of policy space will be updated in one iteration of FP process, and it will result in extremely slow policy update. To remedy this, instead of only considering ``the best path'', we compute top k-best paths at each step, and policy update is done using a (weighted) average over k-best response paths. In experiments we used k as 500 (there are around 20,000 agents in the system). Even with k-best responses, best response paths do not provide enough information to extract a complete policy out of it. Thus, we explore policy completion approaches. 

\subsection{Policy Completion}

Policy completion can be treated as missing data imputation problem. The "right" methods for handling missing data depend strongly on why data is missed. Therefore, it is important to describe the different types of missing data and more specifically the type of missing data in the policy completion problem. 

\citet{rubin1976inference} classified missing data problems into three categories: i) MCAR: where data is missing due to unrelated reason, e.g., due to not collecting required information or collecting incomplete or false information. ii) MAR: where the missing data may only
depend on the observable variables. iii) MNAR: where missing data depends on the values of other variables and its own value. Formally, if $X = (X_{obs}, X_{mis})$ is a data matrix where $X_{obs}, X_{mis}$ represents observed and missing values respectively. $M$ represents missing data pattern. Then,
\begin{flalign}
	& \textbf{MCAR}: P(M|X) = P(M), \qquad \forall X \nonumber \\
	& \textbf{MAR}: P(M|X) = P(M|X_{obs}), \qquad \forall X_{mis} \nonumber \\
	& \textbf{MNAR}: P(M|X) = P(M|X_{obs},Y_{mis}), \quad \exists Y_{mis} \in X_{mis} \nonumber
\end{flalign}

In best response policy, policy value depends on time(t) and state(s) information along with number of hours served(m) and breaks taken(b), since
$$\sum_a \pi^t(s,m,b,a)= 1 \quad \forall t,s,m,b$$
Therefore missing value for an action is also affected by other actions, which may be missing as well. Therefore in this case, missing values in policy belong to the MNAR category. It is shown that non-parametric ``missForest''
~that is based on the unsupervised imputation is robust for MNAR category of missing data~\citep{Guo2021}. 

We now describe the different imputation methods that complete an incomplete policy matrix as input, where incomplete values are usually marked with NAN or NULL. 

\subsubsection{Matrix Factorization}
Matrix factorization (MF)~\citep{lee1999learning} is a strategy for imputing missing values in a matrix. MF takes a partially observed $n \times m$ matrix $X$ as input. $O$ denotes the set of indices in $X$ whose values are known. The goal of MF is to impute missing values in $X (i.e., (i, j) \notin O)$. MF assumes that if $X$, were it fully-known, would be effectively low-rank. That is to say, $X$ can be well approximated by a matrix $Y=U^T V$ of low-rank $r$ ($r$ is a hyper-parameter), where $r \ll min (m,n)$, $U$ and $V$ are matrices with r rows, and $u_i$ and $v_j$ are latent factors that represents $i^{th}$ row and $j^{th}$ column of matrix X. Then $U$ and $V$ can be estimated using following optimization:
$$U,V = \argmin_{U^*,V^*} \sum_{i,j \in O} (x_{ij} - u_i^{T*}v_j^*)^2 + \lambda (||U^*||^2_F + ||V^*||^2_F)$$
where, the second term is the regularization term used to reduce over-fitting. To impute the missing values in $X$, it just needs to compute the inner-product, $u_i^T v_j$ of the learned latent factors.


\subsubsection{MissForest} 
MissForest is one of the best imputation techniques, and is considered as state of art technique for missing data imputation. It is based on random forest imputation algorithm for missing data. This imputation technique works as follows:
\begin{enumerate}
	\item Initialization: Impute all missing data using the mean (for continuous variables) or mode (for categorical variables) value in each data column.
	\item Multiple Imputation Step: 
	\begin{itemize}
		\item Impute: For each variable with missing values, fit a random forest on the observed part and predicts the missing part. 
		\item Stop: Repeat this process of training and predicting the missing values until a stopping criterion is met, or a maximum number of iterations is reached.
	\end{itemize}
\end{enumerate}

MissForest can be applied to mixed data types without any need for pre-processing (no standardization, normalization, scaling, data splitting, etc) on the dataset. It has been proven robust against noises in dataset, due to effective build-in feature selection of random forests. In MissForest, Imputation time, increases with the number of observations, predictors and number of predictors containing missing values. It is based on random forest therefore it inherits the lack of interpretability usually seen in random forests techniques.

MissForest has provided better imputation results in several studies as compared to standard imputation
methods~\cite{Ramosaj2019PredictingMV, tang2017random, hong2020accuracy}. In study conducted by~\cite{waljee2013comparison} missForest was found to be consistently produce the lowest imputation
error when compared against other state of art imputation methods such as k-nearest neighbors (k-NN) imputation and ``MICE''~\cite{van2011mice}. Apart from missForest imputation method, we also experimented with other state of art imputation methods for policy completion. 

\subsubsection{Supervised Imputation (Algorithm~\ref{alg:Supervised})}
Here we provide a supervised imputation method similar to missForest~(Algorithm\ref{alg:Supervised}), where we replace the time consuming random forest regression model with the ground-truth policy. The ground truth policy evolves as planning process goes through the iterations of the FP process. In our experiments we used the average policy learned through fictitious play process as ground-truth policy. As we do not need to train an regression model here, this method is many fold faster than missForest imputation method, while providing comparable results.

\begin{algorithm}[hbt!]
	\caption{Supervised Imputation }\label{alg:Supervised}
	\begin{algorithmic}[1]
		\Require X as n x p matrix
		\State k $\leftarrow$ vector of sorted indices of columns in X
		w.r.t. increasing amount of missing values;
		\State $X^{imp}_{old} \leftarrow$ store previously imputed matrix;
		\For {s in k} do
		\State extract $y^{(s)}_{mis}$ from ground-truth policy using $x^{(s)}_{mis}$;
		\State $X^{imp}_{new} \leftarrow$ update imputed matrix, using predicted $y^{(s)}_{mis}$;
		\EndFor
		\State \textbf{return} the imputed matrix $X^{imp}$
	\end{algorithmic}
\end{algorithm}  

\subsubsection{Multivariate imputation by chained equations (MICE)}
MICE operates under the assumption that given the variables used in the imputation procedure, the missing data are Missing At Random (MAR). Implementing MICE when data are not MAR could result in biased estimates. MICE runs a series of regression models to model each variable with missing data conditioned upon the other variables in the data. This means each variable can be modeled according to its distribution. It works as follows:
\begin{enumerate}
	\item Perform a simple imputation for every missing value, e.g., imputing the mean, treat is as ``place holders''.
	\item Set the ``place holder'' imputed value for one variable (``$v_{miss}$'') back to missing.
	\item Treat ``$v_{miss}$'' as dependent variable and all the other variables as independent variable in a regression model. Regress the observed values from the variable ``$v_{miss}$'' in previous step based on the other variables in the imputation model, which may or may not contain all variables in the dataset.
	\item Replace the missing values for ``$v_{miss}$'' with predictions from the regression model. In subsequent steps, when ``$v_{miss}$'' is used as an independent variable in the regression models for other variables, both observed as well as imputed values will be used.
	\item Repeat steps 2–4 for each variable with a 
	missing value. Once every missing value is imputed, it is considered as one cycle. At the end of a cycle, replace all of the missing values with the predictions from regression model. Repeat it for multiple cycles.
\end{enumerate}

\subsubsection{Generative Adversarial Imputation Nets (GAIN)}
It is relatively new approach to use generative adversarial networks (GANs) for handling missing data. GANs can learn the hidden data distribution very well and through feedback loop between Generator and Discriminator it can result into high accuracy outputs, and therefore it can provide good results for missing data imputation. 

Generative Adversarial Imputation Nets (GAIN) is a very popular GAN architecture for missing data imputation. The idea behind GAIN is very simple: 1) Generator takes the real data as input and imputes the missing values in it. 2) Discriminator takes the imputed data, and tries to figure out which data is imputed.

Generator in GAIN usage same parametric function as any generator (in a GAN network), but it takes more inputs, so that it can fill the missing values close to the true distribution of the data. It generates and uses three different matrices, these matrices are generated from  original incomplete dataset. a)The first matrix (X) is the incomplete data matrix that contains all observed data values and 0 for the missing values; b) Second matrix is a indicator matrix for missing values. i.e., this matrix indicates which of the components of X are being observed (indicated by 1) and which are missing (indicated by 0); c) The last matrix is a random noise matrix that has 0 at the positions of the observed values and random values at the position of missing values.

Discriminator in GAIN tries to distinguish between observed values and imputed value for every data point in $\hat{X}$ (imputed X). It takes $\hat{X}$ as well as a hint matrix as input. The output of Discriminator is a mask matrix that gives the probability of each data point in $\hat{X}$ being observed. The hint matrix provides Discriminator some of the imputed values and some observed values, and therefore it provides support to Discriminator in distinguishing whether a value is imputed or observed.  

\section{Experimental Results}



In this section we provide comparison/analysis of our simulation based best response (SBD) approach equipped with the five imputation methods against an existing approach for strategic planning~\citep{kumar2021adaptive}. We first provide the details of data set and the evaluation method used in this work. Them we provide comparison results for solution quality and run-time of our approaches against the benchmark approach. Next, we analyzed the results to identify best policy for different scenarios. At the end we provide robustness analysis for our different approaches. The comparison/analysis is provided on a real world taxi data set of a company that operates in Singapore. 

\subsection{Dataset Description}
In this work, we use a data set for taxis operating in Singapore. We use 3 months of trip data from 2017. The data are generated from approximately more than 20,000 taxis operating everyday. The data set contains raw information about: a) the pickup/drop-off date and timestamp; b) the taxi trip duration; c) the pickup/drop-off coordinates; d) and the start and end zone ids (entire region is divided in 100 zones). Each pair of a pickup and a drop-off point is defined as a taxi trip. After some data filtering, where we removed taxis operating for more 15 hours in a day, total trips correspond to, a total of more than 0.3 million taxi trips every day.

\subsection{Evaluation method}
To evaluate our method against the benchmark approach~\citep{kumar2021adaptive}, we generate policy for each day of training data set and evaluate it on individual days of test data set.

\textbf{Generate policy for each day:} Using a single day of data, we generated policies using different methods: a) Benchmark method \citep{kumar2021adaptive}; b) Using our SBR method, where SBR is coupled with different imputation techniques studied in this work. An agent can take up to 2 breaks. Policy convergence criteria and maximum number of hours an agent can operate is same across all methods including benchmark approach. We generated such policies for each day in data set. 

\textbf{Training/Testing data set:} For all evaluations/analysis done in this work, 1 week of data is treated as training data set and remaining 3 weeks of the same month is treated as testing data set. All results included in the paper are for week 1 as training set and individual days in weeks 2-4 as test set for the month of March and April (unless stated otherwise). We obtained similar results for other combinations of training and test data.

\subsection{Quality comparison}
\begin{figure*}[hbt!]
	\includegraphics[page=3, width=3.45in, height=2.05in]{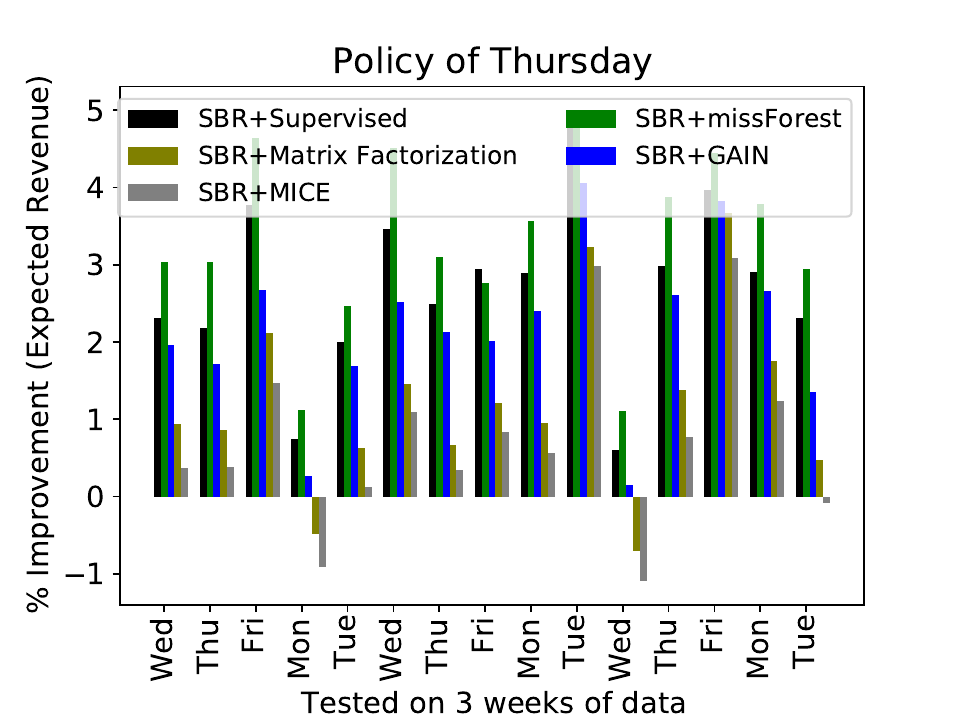}
	\includegraphics[page=7,width=3.45in, height=2.05in]{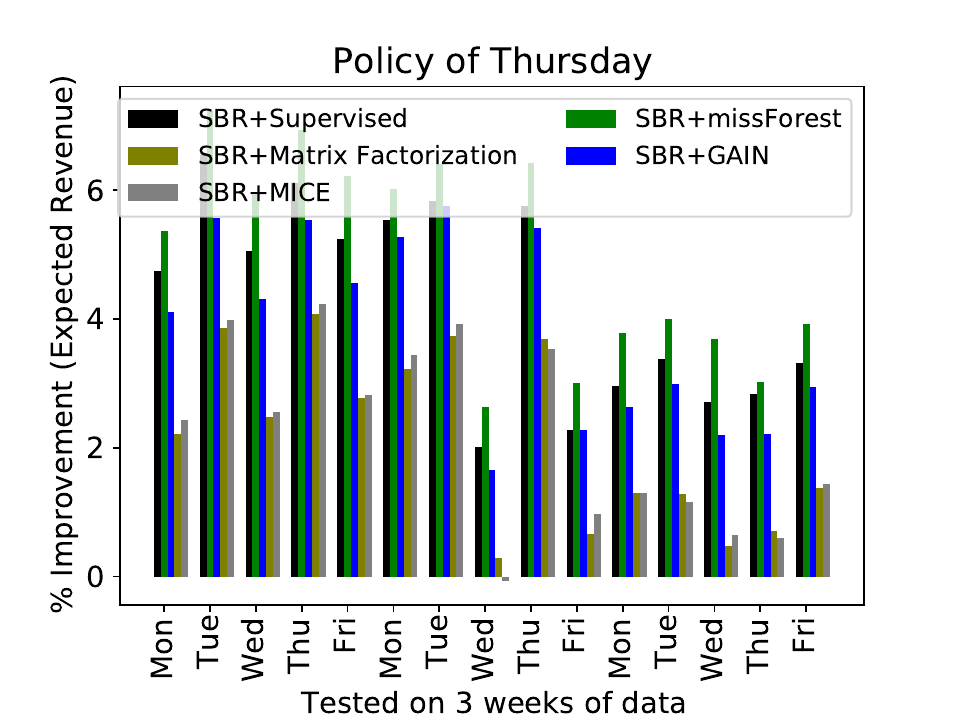}
	\caption{Solution Quality: percentage improvement over benchmark, when tested over 3 test weeks for the policy generated by our simulation based best response method using demand data for Thursday in week 1}
	\label{fig:Solution}
\end{figure*}
\begin{figure*}[hbt!]
	\includegraphics[page=13, width=3.45in, height=2.05in]{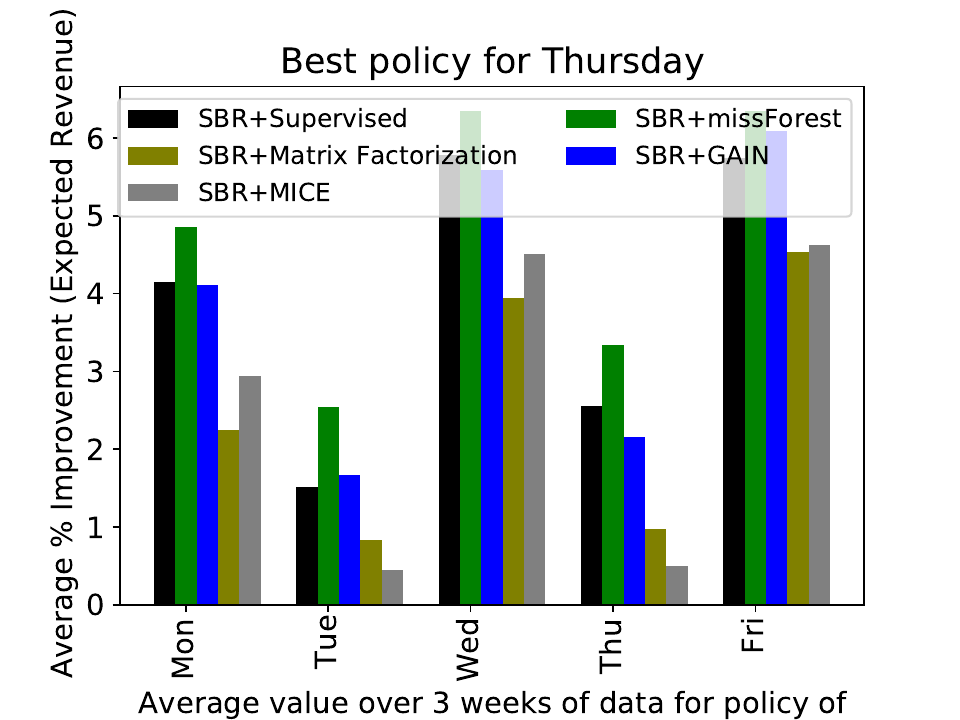}
	\includegraphics[page=13,width=3.45in, height=2.05in]{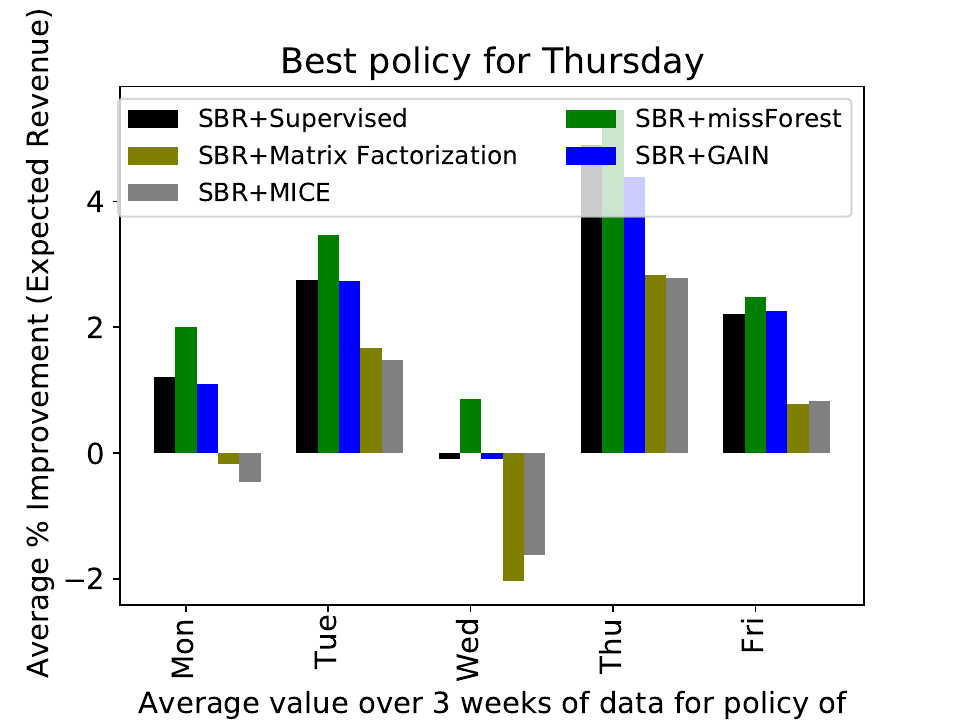}
	\caption{Average Solution Quality (\% improvement over baseline) when different weekday policies are tested on Thursday of 3 test weeks on the month of March and Apr}
	\label{fig:Average}
\end{figure*}

\textbf{Solution quality comparison:} We compare the solution quality (expected revenue) of different methods on test data. Weekday polices are evaluated on weekdays, and weekend policies are evaluated on weekends of the 3 testing weeks.

\textbf{Run-Time comparison:} Here we provide an estimate of (on average) run-time taken by different method to generate a converged policy on training data.

\textbf{Results of Quality Comparison:} We generated 7 polices for each day of training data (week 1) with 5 imputation methods over simulation-based best response and the benchmark approach. 
Key results are as follows:
\begin{itemize}
	\item Our simulation based best response method enabled with missForest Imputation method provides the best results with up to $\approx 10.5\%$ improvement over the benchmark policy. This is followed closely by GAIN and supervised imputation methods.
	\item Our simulation method enabled with matrix factorization and MICE imputation do not fare as well, but are mostly better than the benchmark approach.
	\item Our simulation method enabled with supervised imputation has the least run-time with policies being generated in $\approx$ 10 minutes. In contrast, benchmark approach takes up to 3 days. MICE, missForest, matrix factorization,  and GAIN generate solutions in 3-8 hours, with GAIN being the slowest. We used 100 training epochs for training the GAIN.
\end{itemize}



\subsection{Analysis} 

\textbf{Analyze best policy for different days:} Next, we analyzed the results to identify best policy for different scenarios. 
\begin{itemize}
	\item Best policy for each day (on average): 
	Based on average performance on test set, we can identify best policy for each day, Table~\ref{tab:bestPolicy} shows the best policy for each day of the week with training data as Week1 (March). Similar results are observed with different settings of training and test data.
	\item Best policy for each weekdays/weekend: With week1 (March) as training data, policy generated for Wednesday works well on all weekdays and policy generated for Sunday works well on all weekends. Similar results are observed with different data (different month and week) where one policy works well on all weekdays and one policy works well on all weekends.
\end{itemize}  

\begin{table}[hbt!]
	\begin{tabular}{|p{1.5cm}| p{0.55cm}| c |c| c| c| c| c|} 
		\hline
		\textbf{Day} & Mon & Tues & Wed & Thurs & Fri & Sat & Sun \\ \hline 
		\textbf{BestPolicy} & Fri & Wed & Wed & Fri & Wed & Sun & Sun \\ \hline
	\end{tabular}
	\caption{Best policy for each day (based on avg. performance on 3 test weeks), when trained w/ week 1 data in March.}
	\label{tab:bestPolicy}
\end{table}

\textbf{Robustness analysis:} Here we provide robustness analysis of our approaches. 
We ran the worst case analysis of different polices on 3 test weeks. Detailed result for worst case, best case and average case scenarios with all imputation methods is shown in Table~\ref{tab:Robust_w1_March_Weekday} for training week 1 (March) when tested on weekdays of 3 test weeks, Table~\ref{tab:Robust_w1_March_Weekend} for training week 1 (March) when tested on weekends of 3 test weeks, Table~\ref{tab:Robust_w1_Apr_Weekend} for training week 1 (Apr) when tested on weekdays of 3 test weeks and Table~\ref{tab:Robust_w1_Apr_Weekdays} for training week 1 (Apr) when tested on weekends of 3 test weeks.
In summary, worst case performance of missForest imputation with best policy (one policy for all weekdays/weekend of test days) is as follows, It is providing average improvement of 4.47\% and 3.48\% for weekdays and weekends in March, and 2.48\% and 3.62\% for weekdays and weekends in April.


\begin{table}[H]
	\begin{tabular}{|c |c| c| a |} 
		\hline
		\textbf{Method} & Average Case & Best Case & Worst Case \\ \hline 
		\textbf{Supervised} & 3.86 & 10.67 & -0.22 \\ \hline 
		\textbf{Matrix} & 2.5 & 9.25 & -1.57 \\ \hline 
		\textbf{MICE} & 2.59 & 9.81 & -1.13 \\ \hline 
		\rowcolor{Gray} \textbf{missForest} & 4.47 & 10.57 & 0.58\\ \hline 
		\textbf{GAIN} & 3.84 & 10.93 & -0.01 \\ \hline 
	\end{tabular}
	\caption{[Training data: Week 1 (March)]: Average, Best and Worst Case Performance (\% improvement over baseline) when tested on weekdays of 3 test weeks}
	\label{tab:Robust_w1_March_Weekday}
\end{table}

\begin{table}[H]
	\begin{tabular}{|c |c| c| a |} 
		\hline
		\textbf{Method} & Average Case & Best Case & Worst Case \\ \hline 
		\textbf{Supervised} & 2.49 & 3.99 & 0.15 \\ \hline 
		\textbf{Matrix} & 0.04 & 1.47 & -3.71 \\ \hline 
		\textbf{MICE} & 0.0 & 1.4 & -3.63 \\ \hline 
		\rowcolor{Gray} \textbf{missForest} & 3.48 & 5.04 & 1.7 \\ \hline 
		\textbf{GAIN} & 1.76 & 3.04 & -0.85 \\ \hline 
	\end{tabular}
	\caption{[Training data: Week 1 (March)]: Average, Best and Worst Case Performance (\% improvement over baseline) when tested on weekend of 3 test weeks}
	\label{tab:Robust_w1_March_Weekend}
\end{table}

\begin{table}[H]
	\begin{tabular}{|c |c| c| a |} 
		\hline
		\textbf{Method} & Average Case & Best Case & Worst Case \\ \hline 
		\textbf{Supervised} & 1.8 & 6.6 & -2.72\\ \hline 
		\textbf{Matrix} & 0.16 & 5.06 & -4.55 \\ \hline
		\textbf{MICE} & 0.15 & 4.73 & -4.14 \\ \hline 
		\rowcolor{Gray} \textbf{missForest} & 2.48 & 7.24 & -1.57 \\ \hline 
		\textbf{GAIN} & 1.72 & 6.78 & -2.34 \\ \hline 
		
	\end{tabular}
	\caption{[Training data: Week 1 (Apr)]: Average, Best and Worst Case Performance (\% improvement over baseline) when tested on weekdays of 3 test weeks}
	\label{tab:Robust_w1_Apr_Weekdays}
\end{table}

\begin{table}[H]
	\begin{tabular}{|c |c| c| a |} 
		\hline
		\textbf{Method} & Average Case & Best Case & Worst Case \\ \hline 
		\textbf{Supervised} & 2.25 & 6.41 & -2.21 \\ \hline 
		\textbf{Matrix} & -0.76 & 3.08 & -5.67 \\ \hline
		\textbf{MICE} & -0.4 & 3.22 & -4.91 \\ \hline 
		\rowcolor{Gray} \textbf{missForest} & 3.62 & 7.77 & -1.38 \\ \hline 
		\textbf{GAIN} & 1.11 & 5.76 & -4.73 \\ \hline 
		
	\end{tabular}
	\caption{[Training data: Week 1 (Apr)]: Average, Best and Worst Case Performance (\% improvement over baseline) when tested on weekend of 3 test weeks}
	\label{tab:Robust_w1_Apr_Weekend}
\end{table}

\begin{figure}
    \subfloat[]{\includegraphics[page=2, width=1.6in]{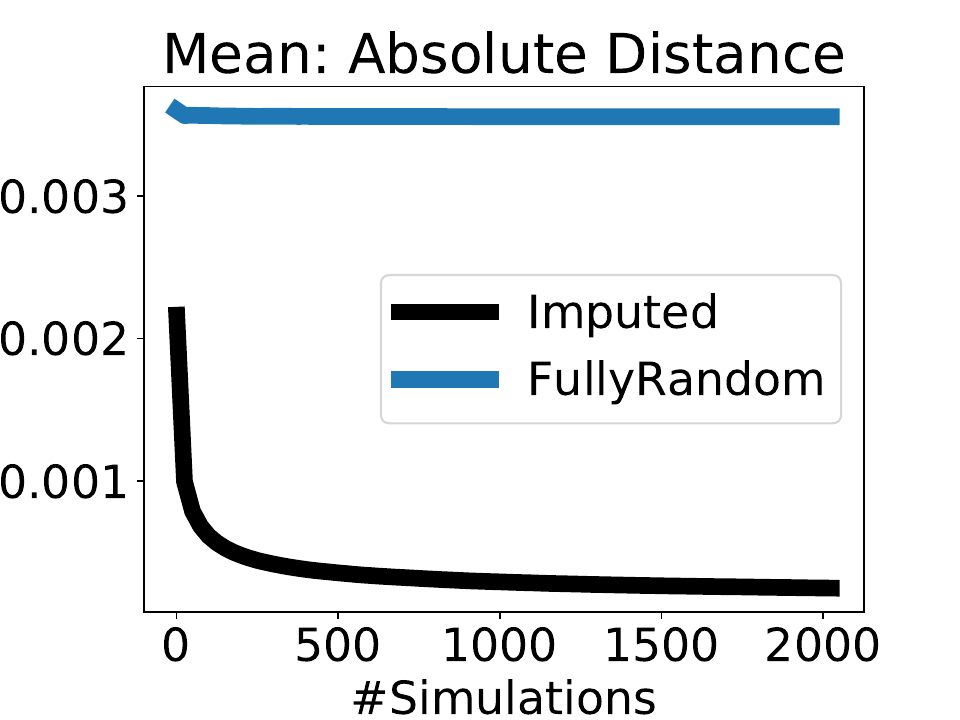} }
    \subfloat[]{\includegraphics[page=4, width=1.6in]{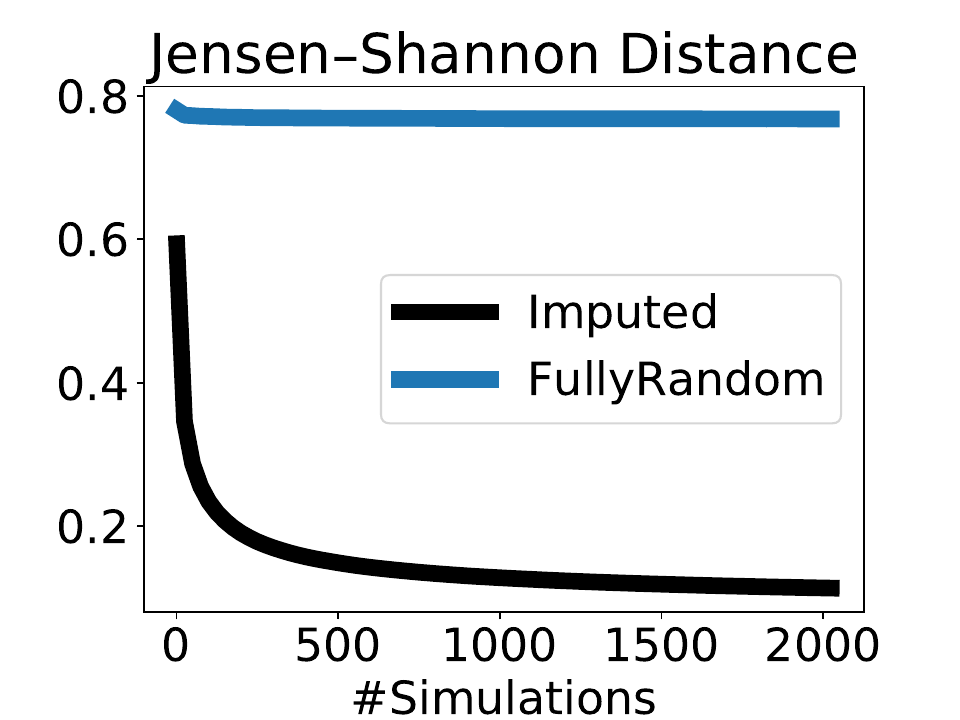} }
    \caption{Distance of occupancy matrix from best response (without imputation) occupancy matrix}
	\label{fig:distance}
\end{figure}

\section{Discussion}
When generating policy from a given set of  trajectories~\citep{ho2016generative}, a theoretical sound way of evaluating correctness is the distance between occupancy measures (visitation frequencies of different state, action pairs) of the policy and that of the trajectories. In Figure~\ref{fig:distance}, we plot  the distance as number of simulations is increased for two measures: mean absolute distance and Jensen-Shannon distance. 
  
We compare the distance of best response policy from (a) our imputation method; and (b) a fully random policy.  For this, we simulated each policy multiple times and compared distance between occupancy matrices (visitation frequencies of different state, action pairs). Occupancy matrix generated by our imputation method policy is much closer to the best response trajectories than the fully random policy. As number of simulations is increased, distance between occupancy matrix generated by imputed policy and the best response policy decreases. 

\section{Conclusion}
In this work we introduce mechanism to provide better flexibility for selfish agent while improving the performance of the entire system.  We introduced a simulation based (faster) equilibrium computation method. We studied and analyzed different imputation methods and show that a good imputation method coupled with a well designed simulation based best response computation can help in achieving better symmetric equilibrium for large scale systems, in a time efficient manner. In our experiments we also show that we can reuse policy for multiple days. We analyzed and provided, (on average) best policy for each day of the week on data set used in this experiment.

\section{Acknowledgment}
This research/project is supported by the National Research Foundation Singapore and DSO National Laboratories under the AI Singapore Programme (AISG Award No: AISG2-RP-2020-016).

\balance
\bibliographystyle{ACM-Reference-Format}
\bibliography{sample}



\end{document}